# Local Mechanical Response of Lipid Membranes to Tilt Deformation

**Konstantin V. Pinigin**

A. N. Frumkin Institute of Physical Chemistry and Electrochemistry, Russian Academy of Sciences, 31/4 Leninskiy Prospekt, 119071 Moscow, Russia

Correspondence: piniginkv@gmail.com

**Abstract:** Using molecular dynamics, this study investigates the local elastic properties of transverse shear deformation of lipid membranes. The analysis demonstrates that transverse shear deformation induces anisotropy in the local stress profile of the lipid bilayer, a phenomenon attributed to the Poynting effect. By analyzing the relationship between transverse shear stress and the induced anisotropy, the local transverse shear modulus is determined. From the local transverse shear modulus, several integral elastic parameters can be derived, including the monolayer tilt modulus, tilt–curvature coupling modulus, and curvature–gradient modulus. The calculated tilt modulus values show good agreement with results from an independent analysis of lipid director fluctuations.

## 1. Introduction

Lipid membranes, ubiquitous in all living organisms, serve as the fundamental organizational principle of cells, demarcating their boundaries and compartmentalizing internal structures [1,2]. Beyond their role as selective barriers that regulate molecular transport, these dynamic assemblies are intimately involved in a myriad of vital cellular processes, including cell division, fusion, endocytosis, and mechanotransduction. Many of these crucial functions necessitate significant morphological transformations of the membrane. Consequently, the deformations of lipid membranes are not merely passive responses to external forces but are often actively controlled and energetically regulated events central to cellular function. These complex mechanical behaviors, ranging from subtle undulations to dramatic shape changes, are typically successfully described within the framework of elasticity theory [3–7]. This theoretical approach provides a powerful means to quantify the energy cost associated with various membrane deformations, offering insights into the underlying physical principles governing their dynamic roles in biological systems.

From the perspective of classical 3D elasticity theory, a lipid monolayer can be conceptualized as a three-dimensional elastic continuum that exhibits fluidity in the lateral dimension [8–11]. This body is characterized by intrinsic nonzero prestress and three independent local elastic moduli: the stretching modulus, the transverse shear modulus, and the lateral shear modulus [9–11]. Experimentally, only integral characteristics derived from these moduli—such as the bending modulus [6] or spontaneous curvature [12] — have been measured. These integral parameters correspond to specific averages over the monolayer thickness.

By contrast, molecular dynamics (MD) simulations offer a distinct advantage: they enable the investigation not only of such integral parameters but also of the local elastic moduli themselves by probing the stress response to applied strains. While the local stretching modulus has been the most extensively studied [13,14] and its values determined in MD simulations [13], the precise values and measurement methodologies for the other two local moduli—the local transverse shear modulus and the local lateral shear modulus—remain largely elusive. Although some efforts have been made to characterize the properties of the local lateral shear modulus [15], its numerical value has not yet been definitively measured. Critically, the characteristics of the local transverse shear modulus have received virtually no attention.

The development of methods to measure the local transverse shear modulus, even if initially only through molecular dynamics (MD) simulations, would provide valuable insights into membrane mechanics. Transverse shear deformation corresponds to lipid tilt, i.e., a deviation of lipid orientation from the membrane normal [9–11,16,17]. The elastic properties associated with this tilt deformation are important for a range of membrane-related phenomena, including membrane-mediated protein interactions [18–22], the phase boundary energy of liquid-ordered domains [23,24], protein-domain interactions [25–28], membrane–protein interactions [19,29], poration [30,31], and membrane fusion and fission [32–35].

Knowledge of the local transverse shear modulus would also enable the determination of other key integral elastic parameters, including not only the tilt modulus but also the tilt–curvature coupling modulus and the curvature–gradient modulus [11]. As demonstrated in Ref. [11], incorporating these moduli into the elastic energy functional can significantly alter the theoretically predicted nature of membrane-mediated interactions between embedded inclusions. Moreover, conventional fluctuation-based methods are insufficient for quantifying the tilt–curvature coupling and curvature–gradient moduli. As shown in Ref. [11], such approaches face methodological limitations, with their results at short wavelengths—precisely the regime relevant to these moduli—failing to match theoretical predictions.

Beyond its role in enabling the determination of complex integral moduli, the local transverse shear modulus is fundamentally important for understanding the mechanical properties of membranes at the nanoscale. Insight into this parameter is essential for elucidating the intricate mechanical interplay between membrane proteins and the surrounding lipid bilayer, providing an elastic continuum framework to describe local deformations induced by protein–membrane interactions.

The aim of this work is the development of an MD method for determining the local transverse shear modulus. Section 2 (Theory) provides a theoretical analysis of transverse shear deformation and demonstrates how this modulus can be measured using MD simulations. Section 3 (Methods) describes the simulation setup and analysis procedures. Results are presented in Section 4 and discussed in Section 5.

## 2. Theory

Consider a single lipid monolayer treated as a continuous elastic body. In the reference configuration, the monolayer is perfectly flat, and the area per lipid corresponds to its equilibrium value. A Cartesian coordinate system is defined to describe the geometry and deformation of the monolayer. The $z$-axis is perpendicular to the monolayer surface, pointing from the lipid headgroups toward the tails. For visualization, the monolayer can be imagined lying horizontally with the lipid tails facing upward, the $z$-axis pointing vertically, the $x$-axis extending to the right, and the $y$-axis directed away from the viewer.

An arbitrary deformation of the monolayer can be described by a vector function $\mathbf{R}(x, y, z)$, which maps points from the reference configuration to their deformed positions. Following classical elasticity theory [9,36], the magnitude of these deformations is quantified by the Green-Lagrange strain tensor $\mathbf{U}$, defined as:

$$\mathbf{U} = \frac{1}{2}\left(\nabla\mathbf{R}\cdot(\nabla\mathbf{R})^T - \mathbf{I}\right), \qquad (1)$$

where $\nabla\mathbf{R}$ is the deformation gradient tensor and $\mathbf{I}$ is the identity tensor. The elastic energy of the monolayer is a function of the components of this strain tensor. Considering the transverse symmetry of the monolayer, the elastic energy density is a function only of the following invariants: $u_{xx}+u_{yy}$, $(u_{xx}-u_{yy})^2+4u_{xy}^2$, $u_{xz}^2$, $u_{yz}^2$, and $u_{zz}$. However, assuming the condition of volume incompressibility, which is well-supported by experimental evidence [37–39] and MD simulations [40–42], the number of independent invariants is reduced, and $u_{zz}$ does not need to be explicitly included in the energy function. In the subsequent analysis, tilt deformation will be considered solely along the $x$-axis. This simplification allows for setting $u_{xy}=0$ and $u_{yz}=0$. Consequently, the elastic energy will be considered a function of $u_{xx}+u_{yy}$, $(u_{xx}-u_{yy})^2$ and $u_{xz}^2$. The general expression for the Cauchy stress tensor $\Sigma$ is given by:

$$\Sigma = \nabla\mathbf{R}\frac{\partial W}{\partial \mathbf{U}}(\nabla\mathbf{R})^T - P\mathbf{I}. \qquad (2)$$

Here, $P$ is an indeterminate hydrostatic pressure, which appears as a Lagrange multiplier due to the volume incompressibility constraint, and $\dfrac{\partial W}{\partial \mathbf{U}}$ is the derivative of the elastic energy density ($W = W(u_{xx}, u_{yy}, u_{xz})$, defined per unit reference volume) with respect to the Green-Lagrange strain tensor $\mathbf{U}$.

Consider a transverse shear deformation along the $x$-axis. In this deformation, points within the monolayer experience a relative shift along the $x$-direction, with the displacement magnitude varying as a function of the $z$-coordinate. This deformation is characterized by the deformation gradient vectors $\mathbf{e}_x = \partial \mathbf{R} / \partial x = (1, 0, 0)$ and $\mathbf{e}_y = \partial \mathbf{R} / \partial y = (0, 1, 0)$. The expression for $\mathbf{e}_z = \partial \mathbf{R} / \partial z$ is given by $(2u_{xz}, 0, 1)$, derived from the definition of the strain tensor, the incompressibility condition, and the perpendicularity of $\mathbf{e}_z$ to $\mathbf{e}_y$. In general, $u_{xz}$ can depend on $z$, meaning its value may vary with depth along the $z$-axis. Substituting these expressions into the equation for the Cauchy stress tensor yields the following result:

$$\Sigma_{xx} = \frac{\partial W}{\partial u_{xx}}(0,0,u_{xz}) + 4\frac{\partial W}{\partial u_{xz}}(0,0,u_{xz})u_{xz} - P, \tag{3a}$$

$$\Sigma_{xx} = \frac{\partial W}{\partial u_{yy}}(0,0,u_{xz}) - P, \tag{3b}$$

$$\Sigma_{xz} = \Sigma_{zx} = \frac{\partial W}{\partial u_{xz}}(0,0,u_{xz}), \tag{3c}$$

$$\Sigma_{zz} = -P, \tag{3d}$$

$$\Sigma_{xy} = \Sigma_{yx} = \Sigma_{yz} = \Sigma_{zy} = 0. \tag{3e}$$

As evident from the derived expressions, when $u_{xz}$ is nonzero, the off-diagonal components $\Sigma_{xz}$ and $\Sigma_{zx}$ are also nonzero. The tensor is symmetric which ensures equilibrium of moments within the body. Furthermore, a condition for equilibrium in the absence of external forces is that the divergence of the Cauchy stress tensor must be zero ($\nabla \cdot \Sigma = 0$). From this, it follows that $\Sigma_{zz}$ = const and $\Sigma_{xz} = \Sigma_{zx}$ = const. Therefore, $\Sigma_{zz}$ equals the external ambient pressure. However, concerning $\Sigma_{xz}$ and $\Sigma_{zx}$, since these components must vanish far from the monolayer (as a boundary condition for an isolated monolayer), it implies that in the absence of external forces, $\Sigma_{xz}$ and $\Sigma_{zx}$ must also be zero throughout the monolayer. Consequently, to maintain a nonzero transverse shear stress in an equilibrium state, the application of external forces is required.

During transverse shear deformation, the lateral stress components, $\Sigma_{xx}$ and $\Sigma_{yy}$, become unequal. In elasticity theory, this phenomenon is known as the Poynting effect [36,43–46], which describes the generation of normal stresses in a material subjected to pure shear. Elastic energy dependence on $u_{xx} + u_{yy}$ and $(u_{xx} - u_{yy})^2$ implies $\frac{\partial W}{\partial u_{xx}}(0,0,u_{xz}) = \frac{\partial W}{\partial u_{yy}}(0,0,u_{xz})$. Consequently, the difference between $\Sigma_{xx}$ and $\Sigma_{yy}$ can be written as:

$$\Sigma_{xx} - \Sigma_{yy} = 4\frac{\partial W}{\partial u_{xz}}(0,0,u_{xz})u_{xz} = 4\Sigma_{xz}u_{xz}. \tag{4}$$

$W(0,0,u_{xz})$ can be written as a Taylor series in terms of $u_{xz}^2$:

$$W(0,0,u_{xz}) = 2\lambda_T(z)u_{xz}^2 + O(u_{xz}^4), \tag{5}$$

where $\lambda_T(z)$ is the local transverse shear modulus, which accounts for the monolayer's resistance to shear deformation at a specific depth $z$. The inclusion of the factor of two in this expression for elastic energy is a standard convention [9,11]. This is because $u_{xz}$ equals half the magnitude of the tilt vector, $\mathbf{T}(z)$, i.e., $u_{xz} = T(z)/2$. Therefore, the $u_{xz}^2$ term effectively captures the square of the full tilt magnitude scaled by a factor of one-fourth, which is then compensated by the factor of two for consistency in energy formulations. The tilt vector is defined as $\mathbf{T}(z) = \frac{\mathbf{n}(z)}{\mathbf{n}(z) \cdot \mathbf{N}(z)} - \mathbf{N}(z)$. Here, $\mathbf{N}(z) = [\mathbf{e}_x \times \mathbf{e}_y] / \|[\mathbf{e}_x \times \mathbf{e}_y]\|$ is the local monolayer normal, while $\mathbf{n}(z) \equiv \mathbf{e}_z / |\mathbf{e}_z|$ is the director, a unit vector representing the average orientation of the lipid tails. A common simplification in many formulations is to assume $\mathbf{n}(z)$ remains constant throughout the monolayer's thickness. However, this is not a necessary constraint in the current framework. From Eqs. (3c) and (5) it follows that $u_{xz} = \Sigma_{xz} / (4\lambda_T(z)) + O(\Sigma_{xz}^2)$. Consequently, Eq. (4) can be expressed as:

$$\Sigma_{xz}^2 = \lambda_T(z)(\Sigma_{xx} - \Sigma_{yy}) + O\left((\Sigma_{xx} - \Sigma_{yy})^2\right) \tag{6}$$

This equation shows that the local transverse shear modulus, $\lambda_T(z)$, acts as the linear coefficient describing the dependence of $\Sigma_{xz}^2$ on $\Sigma_{xx} - \Sigma_{yy}$. This relationship will be used to determine $\lambda_T(z)$ from MD simulations, as further detailed in the methods section.

The local transverse shear modulus profile, $\lambda_T(z)$, provides a direct pathway to determine the integral elastic properties of a lipid monolayer. These properties are derived by integrating the elastic energy density across the monolayer thickness. As demonstrated in Refs. [10,11], the transverse shear deformation is influenced not only by the tilt field, but also by the gradient of the monolayer's effective curvature and the gradient of its stretching deformation. The reference surface for defining tilt, curvature and stretching is typically the neutral surface, which is the surface relative to which the bending modulus is minimal. According to Ref. [11], the integrated quadratic elastic energy expression is given by:

$$w_T = \frac{k_t}{2}\mathbf{T}^2 + k_c \mathbf{T} \cdot \nabla \tilde{K} + \frac{k_{gr}}{2}(\nabla \tilde{K})^2 + B\mathbf{T} \cdot \nabla \alpha - k_c (\nabla \alpha)^2 + C(\nabla \alpha) \cdot (\nabla K) \tag{7}$$

where $\nabla$ is the nabla operator, $\alpha$ is the stretching deformation (the relative change in the neutral surface area), and $\tilde{K}$ is the effective curvature ($\tilde{K} = K + \nabla \cdot \mathbf{T} \approx \nabla \cdot \mathbf{n}$, where $K$ is the conventional geometric curvature and $\mathbf{n}$ is the director field). The monolayer integral elastic parameters $k_t$, $k_c$, $k_{gr}$, $B$, and $C$ are functions of $\lambda_T(z)$ and are defined by integrals over the monolayer thickness:

$$k_t = \int_{m_0} dz\, \lambda_T(z) \tag{8a}$$

$$k_c = -\frac{1}{2}\int_{m_0} dz(z - z_0)^2 \lambda_T(z) \tag{8b}$$

$$k_{gr} = \frac{1}{4}\int_{m_0} dz(z - z_0)^4 \lambda_T(z) \tag{8c}$$

$$B = -\int_{m_0} dz\, \lambda_T(z)(z - z_0) \tag{8d}$$

$$C = \frac{1}{2}\int_{m_0} dz\, \lambda_T(z)(z - z_0)^3 \tag{8e}$$

where $z_0$ is the position of the neutral surface in the reference configuration, and $\int_{m_0} dz$ represents integration over the monolayer thickness in that configuration. Thus, knowledge of the local transverse shear modulus profile $\lambda_T(z)$ allows for the calculation of these integral elastic parameters, providing a direct link between the local mechanical properties and the macroscopic elastic behavior of the monolayer.

## 3. Methods

MD simulations were conducted using the coarse-grained (CG) Martini force field [47], specifically its implicit solvent version [48]. A coarse-grained approach was selected over all-atom simulations primarily to facilitate the preliminary testing of the theoretical framework. Coarse-grained models, by reducing the number of degrees of freedom, allow for significantly longer simulation timescales and the exploration of larger system sizes, which are crucial for initial theoretical validation and parameter space exploration. All-atom simulations provide higher resolution but are computationally expensive, making them less suitable for the initial stages of hypothesis testing where broad trends and qualitative behaviors are sought.

The Martini force field was specifically chosen due to its widespread adoption, extensive validation across a variety of lipid systems, and its proven ability to accurately reproduce key structural and dynamic properties of lipid bilayers. The implicit solvent model was employed to further simplify simulations and reduce computational cost, as the presence of explicit solvent is not critical for calculating the local transverse shear modulus, the primary quantity of interest.

The lipid 1-palmitoyl-2-oleoyl-sn-glycero-3-phosphocholine (POPC) was chosen for these simulations. POPC is a widely studied and well-characterized phospholipid, making it a suitable model system for investigating fundamental bilayer properties [49–51]. Its established compatibility with the dry Martini force field and its demonstrated ability to form stable and well-behaved lipid bilayers in coarse-grained simulations, as evidenced by previous studies [48,52], further justified its selection.

### 3.1 MD parameters

MD simulations were conducted using GROMACS [53,54]. A stochastic dynamics integrator [55] was employed with a time step of 0.03 ps. Electrostatic interactions were handled using the reaction-field method with a 1.2 nm cutoff and a dielectric constant of 15. Van der Waals interactions were treated with a force-switch modifier, smoothly decaying the force between 0.9 nm and the 1.2 nm cutoff. A Verlet cutoff scheme [56] with a 1.4 nm neighbor list radius was applied, and temperature was maintained at 300 K with a 4.0 ps time constant (previous studies have shown that Dry Martini POPC remains fluid over a wide temperature range and does not transition to the gel phase [52]). For simulations where pressure coupling was used, the Berendsen barostat [57] was applied with a pressure coupling time constant of 4.0 ps. The compressibility was set to $3 \times 10^{-4}$ bar$^{-1}$ in the *xy* plane and zero in the *z* direction. The specific pressure coupling scheme—either semi-isotropic or surface-tension—is detailed in the context of the individual simulations.

### 3.2 System Setup

The simulated system consisted of 256 lipid molecules, divided equally into two leaflets to form a bilayer. The simulation box had a fixed dimension in the *z*-axis of 10.5 nm. The lateral dimensions of the box (*x* and *y* axes) were kept equal and fixed at the average box size corresponding to zero lateral tension.

To determine this average size, an initial equilibration procedure was performed. A bilayer was first equilibrated for 100 ns under zero pressure using a semi-isotropic coupling scheme. A subsequent production run of approximately 1 $\mu$s was then performed, also at zero pressure, during which 1000 frames were saved. The average lateral box dimension was calculated from these frames. The data was divided into 20 blocks, and a bootstrap analysis with 1000 samples was conducted to determine statistical uncertainty. This analysis yielded an average lateral box dimension of 9.054 ± 0.002 nm.

A single frame from the production run with a lateral dimension closest to this average value (9.05442 nm) was selected. This frame served as the initial configuration for all subsequent simulations.

*3.3 Application of External Forces*

To generate transverse shear deformation, constant external forces were applied to selected beads of POPC lipid molecules using the GROMACS Colvars module [58]. The coarse-grained POPC model contains 12 beads of equal mass, grouped according to their structural position [48]: NC3 (choline) and PO4 (phosphate) represent the head group; GL1 and GL2 (glycerol moiety) correspond to the interfacial region; and the remaining beads form the hydrocarbon tails—C1A, D2A, C3A, C4A for the oleoyl chain, and C1B, C2B, C3B, C4B for the palmitoyl chain. Forces were imposed on the midpoint of the head beads (NC3 and PO4) and on the terminal tail beads (C4A and C4B). Specifically, forces of magnitude $f/2$ directed along the $x$-axis were applied to NC3 and PO4, while equal forces in the opposite direction were applied to C4A and C4B. This arrangement generated a moment **M** on each lipid molecule, expressed as the cross product **M** = **d** × **f**, where **f** is the total force acting on C4A and C4B, and **d** is the vector connecting the midpoint of NC3 and PO4 to the midpoint of C4A and C4B, oriented from head to tail. Conceptually, this procedure is equivalent to treating a lipid molecule as a line segment spanning from the headgroup midpoint to the terminal tail midpoint, with equal and opposite forces applied at its ends. In addition to the forced deformation, thermal fluctuations of the lipid director — defined as the normalized vector **d** — were analyzed in the absence of external perturbation (see Sec. 3.4), enabling direct comparison of the tilt modulus obtained from fluctuation analysis and from stress profiles.

In the upper monolayer, forces of magnitude $f/2$ were applied to the head beads (NC3 and PO4), directed along the positive x-axis. Simultaneously, forces of magnitude $-f/2$ were applied to the terminal tail beads (C4A and C4B), directed along the negative $x$-axis. In the lower monolayer, the force directions were reversed, ensuring a zero net force on each lipid. The applied force magnitudes ($f$) were varied to 0, 2, 4, and 6 kJ mol$^{-1}$ nm$^{-1}$ to explore different degrees of shear deformation. This force application scheme preserves the symmetry of local stress profiles relative to the monolayer's inter-monolayer surface, simplifying the statistical analysis of data.

For each force magnitude ($f$) applied, the system was subjected to a three-step simulation protocol. First, the force amplitude was ramped up linearly over 30 ns to gradually deform the bilayer and avoid large-scale structural disruptions. This was followed by a 100 ns equilibration period to ensure the system reached a stable state under the constant external forces. Finally, a 3 $\mu$s production run was conducted for data collection and analysis.

Alternative force application schemes might be suggested. The force application scheme described above is the one for which the main results will be presented. In the Discussion section, results will also be compared to two alternative force application schemes: i) a force of magnitude $f$ is applied only to the NC3 bead in the head region, and ii) forces of equal amplitude ($f/2$) are applied to the GL1 and GL2 beads of the glycerol moiety. For all schemes, forces applied to the terminal beads of the tails will be the same.

*3.4 Analysis of Stress Profiles*

Local stress profiles were calculated using the GROMACS-LS v2016.3 software [59–62]. This tool implements the Irving–Kirkwood–Noll definition of local stress [63,64], employing spatial averaging over a 3D rectangular grid with trilinear weighting functions. To generate these profiles, simulation frames were saved every 4.98 ps during the production runs. The resulting data was then divided into 20 blocks for subsequent statistical analysis. The grid step for the stress profiles was set to 0.05 nm.

Prior to analysis with GROMACS-LS, the lipid bilayer was centered within the simulation box using MDAnalysis version 2.9.0 [65,66]. The resulting stress profiles were then shifted so that the bilayer center corresponded to $z = 0$. To improve statistical accuracy by leveraging the system's symmetry with respect to the $z$-axis, the stress profiles were symmetrized according to the formula: $S(z) = (S(z) + S(-z))/2$, where $S(z)$ represents any component of the stress tensor.

### 3.4.1 Calculation of Local Transverse Shear Modulus

To determine the local transverse shear modulus, the relationship between the stress tensor components was analyzed. At each $z$-position, the calculated values of $\Sigma_{xz}^2$ and $\Sigma_{xx} - \Sigma_{yy}$ were fitted to a quadratic function constrained to pass through the origin. This constraint reflects the condition that, in the absence of an external force ($f = 0$), both stress components vanish, i.e., $\Sigma_{xz}^2 = \Sigma_{xx} - \Sigma_{yy} = 0$.

A bootstrap method was used to determine the fitting coefficients. For each force value, 1000 resamples were generated by randomly drawing with replacement from the 20 available data blocks. For each resample, the values of $\Sigma_{xz}^2$ and $\Sigma_{xx} - \Sigma_{yy}$ were calculated and fitted to the quadratic function. The linear coefficient of this fit, which corresponds to the local transverse shear modulus according to Eq. (6), was then recorded. The final modulus and its standard deviation were obtained from the distribution of the 1000 bootstrap estimates.

### 3.4.2 Calculation of Integral Elastic Parameters

The local transverse shear modulus, $\lambda_T(z)$, was used to calculate the integral elastic parameters defined by Eqs. (8a–e). Integration was performed using a composite approach, combining Simpson's 1/3 rule with Simpson's 3/8 rule for the final segment. Due to symmetry, the integration was carried out over the lower monolayer, with the upper limit set at the bilayer midsurface. In the implicit solvent model, the stress profiles decay to zero away from the bilayer; therefore, the lower limit was taken at the point where the stress components vanished. These integrations were performed at each step of the previously described bootstrap procedure, yielding 1000 values for each elastic parameter. The mean and standard deviation were subsequently calculated from these values.

To perform these integrations, the position of the neutral surface, $z_0$, was required. The method outlined in Ref. [11] was used to determine $z_0$. This involved simulating the bilayer under various surface tensions, specifically at –100, –50, 0, 75, and 150 bar nm, using the surface-tension pressure coupling scheme. Each of these simulations ran for 3 μs. The resulting local stress profiles were then analyzed according to the methodology of Ref. [11] to determine the value of $z_0$.

### 3.5 Fluctuation Analysis of the Director Field

The tilt modulus was also determined using an alternative method based on the analysis of fluctuations in the lipid director field within the bilayer, without applying external forces. The theoretical framework for this approach is well-established in the literature [10,11,67–71]. However, discrepancies have been noted between theoretical predictions and measured fluctuation spectra in MD simulations [11]. It is generally understood that this divergence is most pronounced at short wavelengths, suggesting an effect at small length scales not fully captured by current theories.

For the purpose of calculating the tilt modulus, it is sufficient to analyze fluctuations at zero wave vector [68,69]. In the fluctuation-based method, one examines fluctuations of the vector $\mathbf{n}_m = (\mathbf{n}_u - \mathbf{n}_l)/2$, where $\mathbf{n}_u$ and $\mathbf{n}_l$ are the in-plane ($xy$) components of the director fields of the upper and lower monolayers, respectively [68–70]. To obtain these quantities, the membrane is divided into a discrete grid, with $\mathbf{n}_u$ and $\mathbf{n}_l$ assigned to each grid cell, followed by a discrete Fourier transform of $\mathbf{n}_m$. In this framework, the first component of $\mathbf{n}_{m0}$ corresponds to the grid-averaged $x$-component of $\mathbf{n}_m$, while the second corresponds to the grid-averaged $y$-component. According to the equipartition theorem, the mean-squared fluctuation of the zero-wave-vector mode is directly related to the monolayer tilt modulus: $\langle \mathbf{n}_{m0}^2 \rangle = k_B T / S k_t$, where $S$ is the membrane area [69].

Whereas conventional fluctuation analysis typically involves discretizing the membrane into a spatial grid and applying a discrete Fourier transform, the present study focuses exclusively on fluctuations at zero wave vector. This approach permits treating the entire membrane as a single unit, effectively corresponding to a 1×1 grid. Accordingly, no spatial grid division was applied in this analysis.

One of the key aspects in analyzing director fluctuations is the normalization of director vectors, which is performed in two stages. First, the director of each lipid is defined, requiring an initial normalization. Second, after averaging the lipid directors within a grid cell, the resulting vector is normalized to assign a single director value to that cell. Two normalization schemes are commonly used: L-normalization and Z-normalization [71]. In L-normalization, the director is scaled to unit length, consistent with continuum theory. In Z-normalization, it is scaled such that its $z$-component equals one, an approach motivated by the Monge gauge parameterization of the membrane [71]. In this study, the results are reported for both L-normalization and Z-normalization, with standard deviations estimated by dividing the trajectory into 20 blocks and applying bootstrap analysis.

## 4. Results
### 4.1 Unperturbed Bilayer Properties

This section presents the results for a free-standing lipid bilayer, unperturbed by external forces. Figure 1 shows the profiles of the local stress tensor components as a function of the coordinate along the bilayer thickness. The only nonzero stress components are the lateral ones, $\Sigma_{xx}$ and $\Sigma_{yy}$. The values of $\Sigma_{xx}$ and $\Sigma_{yy}$ are statistically indistinguishable, confirming the expected lateral isotropy of the bilayer. The normal stress component, $\Sigma_{zz}$, is zero, which is consistent with the implicit solvent model where no external pressure is applied along the $z$-axis.

The lateral stress profile exhibits a characteristic three-peak structure, symmetrically arranged around the bilayer's midsurface. A strong repulsive stress peak is observed in the hydrophobic core at the bilayer center. Moving outwards, an attractive stress well is located in the transition region between the lipid tails and headgroups. Finally, a second repulsive stress peak is present in the hydrophilic headgroup region. This characteristic profile of attractive and repulsive forces across the bilayer thickness is a well-established feature of interactions within lipid membranes [13,48,59].

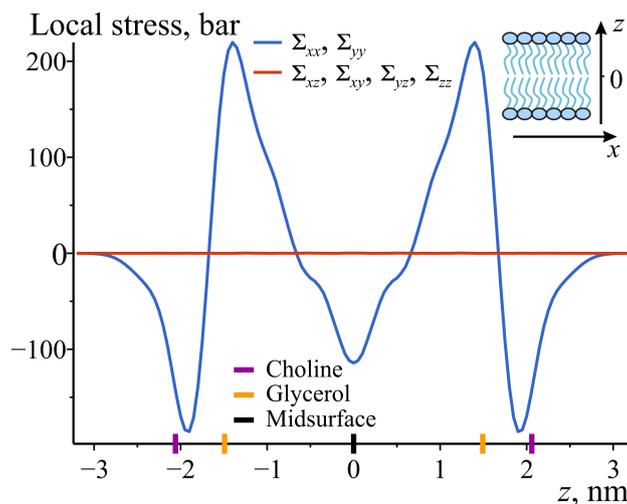

Figure 1. Local stress profiles of an unperturbed lipid bilayer. The inset shows a schematic of the unperturbed planar bilayer with the $z$- and $x$-axes indicated. Average positions of the choline group (purple ticks) and the midpoint between the two glycerol beads (GL1 and GL2, orange ticks) denote the headgroup and interfacial regions, respectively, while the bilayer midsurface is marked by a black tick. Standard deviations are not shown, as they are too small to be resolved graphically.

*4.2 Response to External Forces*

The bilayer's response to constant external forces, described in the Methods section and inducing transverse shear deformation, is now examined. Figure 2 illustrates the results for an applied force magnitude of $f = 4$ kJ mol$^{-1}$ nm$^{-1}$. In contrast to the unperturbed bilayer, a nonzero $\Sigma_{xz}$ component of the local stress tensor appears. Due to the symmetry of the applied forces, this $\Sigma_{xz}$ profile is also symmetric with respect to the bilayer midsurface. Furthermore, the lateral stress components, $\Sigma_{xx}$ and $\Sigma_{yy}$, are no longer equal, with $\Sigma_{xx}$ now being greater than $\Sigma_{yy}$. The remaining stress tensor components — the normal component ($\Sigma_{zz}$) and the other off-diagonal components ($\Sigma_{xy}$, $\Sigma_{yz}$) — remain zero, as expected.

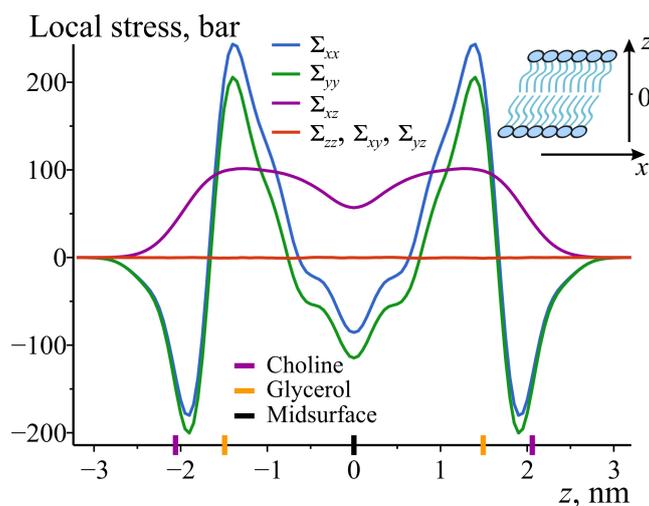

Figure 2. Local stress profiles of a bilayer experiencing transverse shear deformation due to constant external forces ($f = 4$ kJ mol$^{-1}$ nm$^{-1}$). The inset shows a schematic of a planar bilayer subjected to transverse shear, with the $z$- and $x$-axes indicated. Average positions of the choline group (purple ticks) and the midpoint between the two glycerol beads (GL1 and GL2, orange ticks) denote the headgroup and interfacial regions, respectively, while the bilayer midsurface is marked by a black tick. Standard deviations were omitted because their magnitude is too small to be resolved graphically.

As established in the Theory section, the local transverse shear modulus, $\lambda_T(z)$, is the linear coefficient of the relationship between $\Sigma_{xz}^2$ and $\Sigma_{xx} - \Sigma_{yy}$. The determined profile of $\lambda_T(z)$, obtained through the fitting procedure described in the Methods section, is shown in Figure 3. The modulus attains its maximum near the glycerol backbone of the lipid molecules, then gradually decreases with distance from the bilayer, approaching zero in the bulk region. Toward the bilayer midsurface, $\lambda_T(z)$ also decreases, reaching a minimum at the center. Using the determined $\lambda_T(z)$ profile and the previously calculated position of the neutral surface ($z_0 = -1.197 \pm 0.002$ nm, see Methods), the surface integral elastic parameters were calculated according to Eqs. (8a–e). The results of this analysis are as follows: $k_t = 10.4 \pm 0.1\ k_\mathrm{B}T/\mathrm{nm}^2$, $k_c = -1.93 \pm 0.03\ k_\mathrm{B}T$, $k_{gr} = 0.75 \pm 0.03\ k_\mathrm{B}T\ \mathrm{nm}^2$, $B = 0.07 \pm 0.1\ k_\mathrm{B}T/\mathrm{nm}$, $C = 0.1 \pm 0.05\ k_\mathrm{B}T\ \mathrm{nm}$.

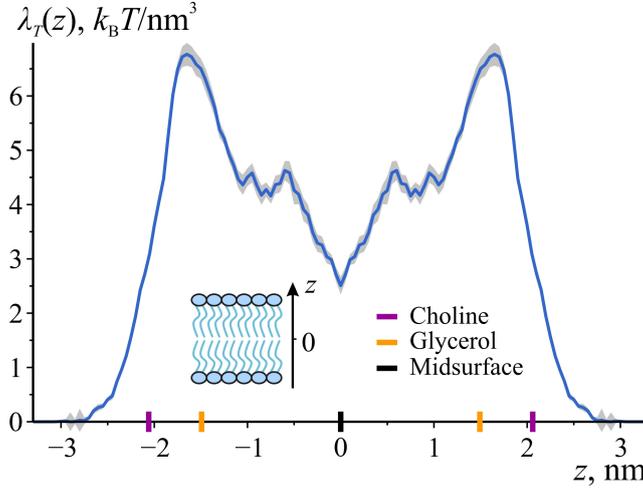

Figure 3. Local transverse shear modulus $\lambda_T(z)$ as a function of the coordinate $z$ along the bilayer normal. The inset shows a schematic of the unperturbed planar bilayer with the $z$-axis indicated. Average positions of the choline group (purple ticks) and the midpoint between the two glycerol beads (GL1 and GL2, orange ticks) denote the headgroup and interfacial regions, respectively, while the bilayer midsurface is marked by a black tick. Gray shading indicates the standard deviation.

*4.3 Fluctuation Analysis of the Director Field*

The tilt modulus was also determined by analyzing the fluctuations of the lipid directors, as described in the methods section. Using L-normalization, the tilt modulus was found to be $10.3 \pm 0.1\ k_\mathrm{B}T/\mathrm{nm}^2$. This value shows excellent agreement with the result of $10.4 \pm 0.1\ k_\mathrm{B}T/\mathrm{nm}^2$ obtained from the stress profile analysis.

In contrast, the use of Z-normalization yielded a significantly lower value of $6.1 \pm 0.5\ k_\mathrm{B}T/\mathrm{nm}^2$. The substantial difference in the result can be attributed to the behavior of the $z$-component of the director vectors. In some frames, the $z$-component becomes very small, causing the normalization procedure to artificially inflate the $n_x$ and $n_y$ components. This inflation leads to an overestimation of the director fluctuations and, consequently, a significant underestimation of the tilt modulus.

## 5. Discussion
*5.1 The Poynting Effect*

MD simulations were performed to investigate transverse shear deformation in lipid membranes. The results show that such deformation induces anisotropy in the lateral local stress (Figure 2). Classical elasticity theory likewise predicts this anisotropy, known as the Poynting effect [36,43–46]. The Poynting effect describes unequal normal stresses arising under shear forces, in addition to the expected shear stresses. It is a characteristic nonlinear phenomenon in elasticity: absent in infinitesimal strain theory, which neglects quadratic terms in the Green–Lagrange strain tensor, but present in finite strain theory, which retains them. In lipid membranes, transverse shear deformation corresponds to lipid tilt, which is the deviation of the lipid director from the local membrane normal [9–11,16,17]. Anisotropy in lateral stress associated with a nonzero tilt field is also predicted by phenomenological two-dimensional models that treat membranes as continuous surfaces [20,72].

The MD data reveal a similar anisotropy. For applied forces of $f = 2$, $f = 4$, and $f = 6$ kJ mol$^{-1}$ nm$^{-1}$, the tension along the $x$-axis (the integral of $\Sigma_{xx}$ over the monolayer thickness) is $17.6 \pm 0.6$, $66 \pm 1$, and $124 \pm 1$ bar nm, respectively. Conversely, the tension along the $y$-axis (the integral of $\Sigma_{yy}$ over the monolayer thickness) is $-20 \pm 1$, $-79.7 \pm 1.2$, and $-172.4 \pm 0.9$ bar nm, respectively. These results show that a positive tension arises along the $x$-axis, while a negative tension arises along the $y$-axis. This distinct behavior confirms the presence of the Poynting effect in lipid membranes under transverse shear.

*5.2 Local Transverse Shear Modulus*

In general, accounting for the condition of volume incompressibility, a lipid monolayer is characterized by three independent local elastic moduli: the stretching modulus, the transverse shear modulus, and the lateral shear modulus [9–11]. While experimental measurement of these moduli remains challenging, molecular simulations offer a powerful alternative for their determination. To date, only the local stretching modulus has been well-studied [13]. The values of the other two moduli have not been previously measured. The properties of the local lateral shear modulus were investigated in Ref. [15], but specific values were not determined.

This study focuses on the local transverse shear modulus, $\lambda_T(z)$, as the least-studied elastic parameter. The measurement of this modulus was made possible by the Poynting effect, which arises during transverse shear deformation. As shown in Eq. (6), the relationship between $\Sigma_{xz}^2$ and $\Sigma_{xx} - \Sigma_{yy}$ under transverse shear deformation along the $x$-axis allows for the calculation of $\lambda_T(z)$. The value of this modulus is presented in Figure 3. A maximum value of approximately 7 $k_BT$/nm$^3$ is reached in the interface region. In the tail region, the modulus decreases to approximately 3 $k_BT$/nm$^3$ at the midsurface.

The local transverse shear modulus, $\lambda_T(z)$, can be directly interpreted as a local tilt modulus, and its biological significance is considerable. This modulus is crucial for how membranes accommodate local tilt deformations that arise from integral membrane proteins and other environmental factors. The spatial profile of $\lambda_T(z)$ reveals a differential mechanical response across the lipid monolayer's thickness, with a significantly stiffer interface region compared to the hydrophobic core. The obtained values of $\lambda_T(z)$ highlight key mechanical properties of the transverse shear deformation that are likely to play a critical role in regulating the conformation and function of membrane-associated molecules, offering new insights into the mechanisms of protein-membrane interactions.

*5.3 Systematic Uncertainty*

Different approaches can be used to apply forces to lipids to achieve transverse shear deformation. The results presented were based on a setup where equal forces were applied to both the head beads and the terminal tail beads. Alternative force application schemes, however, could be considered, leading to different $\lambda_T(z)$ profiles. For instance, in the head region, a force could be applied only to the NC3 head bead (with doubled magnitude to maintain zero net force on the lipid), while the tail forces remain unchanged. Simulations performed with this alternative scheme show that, although the general shape of the $\lambda_T(z)$ profile remains qualitatively similar, the quantitative values differ. The integral elastic parameters obtained from this alternative force application scheme are provided in Table I. A comparison with the first setup shows that the monolayer tilt modulus is lower, and other elastic parameters also differ. This highlights a fundamental systematic uncertainty: the calculated elastic properties depend on how the external forces are applied.

TABLE I. This table summarizes monolayer elastic parameters, derived from different force application methods. The first column identifies the specific beads to which forces were applied along the x-axis in the upper monolayer and in the opposite direction in the lower monolayer. Within the lipid tails, forces of equal magnitude were also applied to the terminal C4A and C4B beads, with a positive x-axis direction in the lower monolayer and a negative x-axis direction in the upper monolayer. Columns 2–5 display elastic parameter values obtained from analyses of stress profiles. The final two columns present the tilt modulus, calculated from director fluctuations using both L-normalization and Z-normalization. The director is defined as a vector from the midpoint of the force-applied beads to the midpoint of the C4A and C4B beads. Standard deviations are provided in parentheses.

| Beads | $k_t$, $k_BT$/nm² | $k_c$, $k_BT$ | $k_{gr}$, $k_BT$ nm² | $B$, $k_BT$/nm | $C$, $k_BT$ nm² | $k_t$, L-norm., $k_BT$/nm² | $k_t$, Z-norm., $k_BT$/nm² |
|---|---|---|---|---|---|---|---|
| NC3, PO4 | 10.4(0.1) | −1.93(0.03) | 0.75(0.03) | 0.07(0.1) | 0.1(0.05) | 10.3(0.1) | 6.1(0.5) |
| NC3 | 8.9(0.2) | −1.62(0.03) | 0.65(0.01) | −0.44(0.07) | 0.22(0.02) | 9.4(0.1) | 3.9(0.5) |
| GL1, GL2 | 7.9(0.3) | −1.11(0.03) | 0.41(0.01) | −1.94(0.09) | 0.81(0.02) | 7.8(0.1) | 7.9(0.1) |

A similar problem exists in fluctuation-based methods, where the results for elastic parameters depend on the specific choice of beads used to define the lipid director or the membrane surface [67–69,71]. In contrast, continuum theory posits a single, unique local transverse shear modulus, $\lambda_T(z)$, which in turn yields a unique tilt modulus and other elastic parameters [9–11]. However, continuum theory remains rather vague about what a lipid director is, simply defining it as the average orientation of lipid molecules in space.

The force application scheme's influence on the results isn't necessarily a flaw. As described in the Methods section, this scheme defines the vector to which the moment is applied. This vector, in turn, can be reasonably considered to define the direction of the lipid director. Consequently, since the vector changes with the force application scheme, the corresponding definition of the director also changes. The results of this study support this view: the tilt modulus of 10.3 ± 0.1 $k_BT$/nm² from the fluctuation method (with L-normalization) agrees well with the 10.4 ± 0.1 $k_BT$/nm² result from the stress profile analysis using the force application scheme described in the methods. Similarly, when the alternative force application scheme is used, the director is redefined as the vector connecting the NC3 bead to the midpoint of the C4A and C4B beads. The fluctuation analysis then yields a tilt modulus of 9.4 ± 0.1 $k_BT$/nm² (with L-normalization), which is in reasonable agreement with the 8.9 ± 0.2 $k_BT$/nm² result from the corresponding stress profile analysis. This suggests that the dependence of $\lambda_T(z)$ on the force application method may be a valid reflection of the director's definition.

In some studies analyzing director fluctuations, the lipid director is defined as pointing from the glycerol region — specifically, from the midpoint of the GL1 and GL2 beads — rather than from the head region [68,69]. This definition corresponds to applying forces to the midpoint of the GL1 and GL2 beads, with forces of equal magnitude ($f$/2) applied to each bead. However, when forces are applied in this manner, the results slightly diverge from theoretical predictions. As shown in Figure 4, a region of negative $\Sigma_{xx} - \Sigma_{yy}$ appears within the lipid heads, while $\Sigma_{xz}$ remains positive in the same area. At points where $|z| \approx 1.8$ nm, $\Sigma_{xx} - \Sigma_{yy} = 0$, yet $\Sigma_{xz}$ is positive. This observation contradicts Eq. (6), which states that if $\Sigma_{xx} - \Sigma_{yy} = 0$, then $\Sigma_{xz}$ must also be zero. At these specific points, $\lambda_T(z)$ is indeterminate. This issue does not arise when forces are applied to the head region, as both $\Sigma_{xx} - \Sigma_{yy}$ and $\Sigma_{xz}$ are non-negative at all points. Although the exact cause of this discrepancy is unknown, the effect is likely minor. The negative values of $\Sigma_{xx} - \Sigma_{yy}$ are relatively small and localized to a narrow region. The point where $\Sigma_{xx} - \Sigma_{yy}$ becomes zero does not depend much on the force magnitude. If the zone where $\Sigma_{xx} - \Sigma_{yy}$ is negative is disregarded and $\lambda_T(z)$ is calculated in the range of $|z| < 1.7$ nm, the resulting tilt modulus value is 7.9 ± 0.3 $k_BT$/nm². The result aligns with values obtained from director fluctuations, specifically 7.8 ± 0.1 $k_BT$/nm² for L-normalization and 7.9 ± 0.1 $k_BT$/nm² for Z-normalization. Other integral elastic parameters for this force application scheme, also calculated within the same range, are provided in Table I.

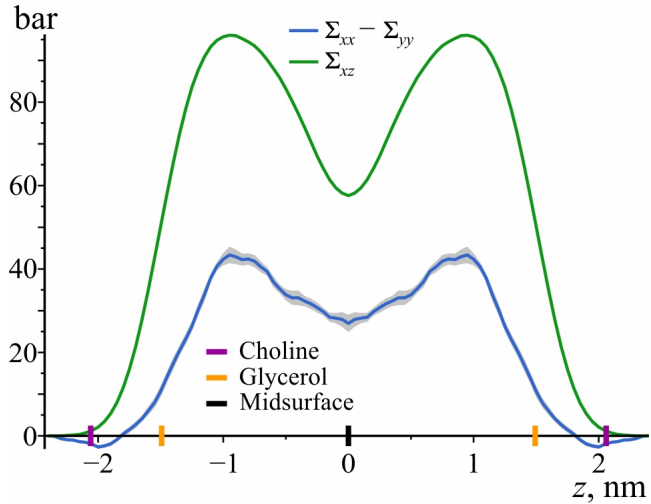

Figure 4 Lateral stress anisotropy ($\Sigma_{xx} - \Sigma_{yy}$) and shear stress ($\Sigma_{xz}$) across the lipid bilayer when external forces ($f$ = 4 kJ mol$^{-1}$ nm$^{-1}$) are applied to the glycerol moiety. Average positions of the choline group (purple ticks) and the midpoint between the two glycerol beads (GL1 and GL2, orange ticks) denote the headgroup and interfacial regions, respectively, while the bilayer midsurface is marked by a black tick. Gray shading indicates the standard deviation for $\Sigma_{xx} - \Sigma_{yy}$. Standard deviations for $\Sigma_{xz}$ are too small to be visually represented.

### 5.4 L-normalization vs. Z-normalization in Fluctuation Analysis

The analysis of director fluctuations yielded a tilt modulus of $10.3 \pm 0.1\ k_BT/\text{nm}^2$ with L-normalization and $6.1 \pm 0.5\ k_BT/\text{nm}^2$ with Z-normalization. The choice between these two normalization schemes represents one of the key systematic uncertainties in fluctuation-based methods [71]. In L-normalization, the director is normalized to a unit length, whereas in Z-normalization, the $z$-component is normalized to a value of one [71].

The value obtained with L-normalization, $10.3 \pm 0.1\ k_BT/\text{nm}^2$, shows excellent agreement with the result of $10.4 \pm 0.1\ k_BT/\text{nm}^2$ from the stress profile analysis. In stark contrast, the value from Z-normalization is significantly lower and has a larger standard error. The primary issue with Z-normalization arises when the $z$-component of a director becomes very small in some frames. This leads to an artificial inflation of the director's $x$ and $y$ components during the normalization process. Although these frames could be excluded, the entire dataset was analyzed to ensure a consistent comparison with the stress profile results. The L-normalization method, in contrast, appears robust to these large fluctuations, with the result remaining very close to the value from the stress profile analysis.

This finding is also supported by the alternative force application scheme, which involved applying forces solely to the NC3 head bead in the head region. In this case, the fluctuation analysis yielded a tilt modulus of $9.4 \pm 0.1\ k_BT/\text{nm}^2$ with L-normalization and $3.9 \pm 0.5\ k_BT/\text{nm}^2$ with Z-normalization. The corresponding value from the stress profile analysis was $8.9 \pm 0.2\ k_BT/\text{nm}^2$. Once again, the L-normalized result is in closer agreement with the value from the stress profiles. Agreement between L- and Z-normalization is observed only when forces are applied to the glycerol moiety (see Table I). These comparisons suggest that L-normalization is a more robust and methodologically sound approach for calculating the tilt modulus from director fluctuations compared to Z-normalization.

### 5.5 Integral Elastic Parameters

In this study, a calculation was performed for not only the tilt modulus but also other integral elastic parameters, which are moments of the $\lambda_T(z)$ profile, as defined by equations (8b–e). These parameters account for the coupling between tilt deformation, membrane curvature, and in-plane strain [10,11,73]. While these parameters were not considered in classical works by Helfrich [4] and later developments by Hamm and Kozlov [9], subsequent studies showed that properly accounting for the dependence of the tilt field on the $z$-coordinate introduces additional quadratic energy terms into the elastic free energy functional [10,11,73]. The theoretical study in Ref. [11] demonstrated that specific values for these additional energy terms could alter the character of membrane-mediated interactions between membrane inclusions, with amphipathic peptides serving as an example. However, the exact values of these additional elastic parameters and their overall influence on lipid membrane mechanics have remained open questions. Experimental measurement of these parameters is challenging, and attempts to derive the parameters from fluctuation analysis on small wavelengths have been problematic due to significant deviations between theory and simulation at these scales [11].

This work provides measurements of these parameters. The calculated values, relative to the neutral surface located at $z_0 = -1.197$ nm within the lipid tails, are: $k_c = -1.93 \pm 0.03\ k_BT$, $k_{gr} = 0.75 \pm 0.03\ k_BT\ \text{nm}^2$, $B = 0.07 \pm 0.1\ k_BT/\text{nm}$, and $C = 0.1 \pm 0.05\ k_BT\ \text{nm}$. Ref. [11] demonstrated that at certain values of these elastic parameters, the nature of membrane-mediated interactions between membrane inclusions can change. As an example, amphipathic peptides were considered. It is worth noting that in Ref. [11], when the membrane-mediated interaction of peptides was considered, the interface surface was used as the reference surface. This is the contact plane between the hydrophilic and hydrophobic parts of the monolayer and is located near the glycerol backbone. It is convenient to define deformations relative to this surface because an amphipathic peptide adsorbed to the membrane is typically situated in this region, exposing the peptide's hydrophilic and hydrophobic parts to the corresponding membrane regions. When re-calculated relative to this interface surface, the parameters are: $k_c = -2.3 \pm 0.04\ k_BT$, $k_{gr} = 1.24 \pm 0.03\ k_BT$ nm², $B = -2.75 \pm 0.1\ k_BT$/nm, and $C = 1.78 \pm 0.06\ k_BT$ nm. The most important of these parameters are $k_c$ and $k_{gr}$, as the parameters $B$ and $C$ describe the coupling of tilt to in-plane strain, which is generally a stiff deformation mode and thus relatively small. Membrane-mediated interactions between peptides are such that as two peptides approach each other, the peptides' local membrane deformations overlap. At very short distances, the elastic deformation energy in the region between the peptides increases sharply, leading to a repulsive force [11,74,75], but at an intermediate distance of approximately 4 nm, the Hamm-Kozlov functional predicts a local energy minimum [11,74,75]. This minimum is hypothesized to facilitate the lateral association of peptides, which may lead to the formation of a pore in the membrane. As demonstrated in Ref. [11], the depth of this local energy minimum can be reduced depending on the values of $k_c$ and $k_{gr}$. However, this reduction is significant only for large absolute values of these parameters. The local energy minimum disappears entirely at values of $k_c = -11.5\ k_BT$ and $k_{gr} = 12\ k_BT$ nm² [11]. The values obtained in this work ($k_c = -2.3 \pm 0.04\ k_BT$ and $k_{gr} = 1.24 \pm 0.03\ k_BT$ nm²) are far from these critical values. At these smaller magnitudes, the Hamm-Kozlov functional is modified only quantitatively and to an insignificant extent. With regard to alternative force application schemes, the values for $k_c$ and $k_{gr}$ do not depend significantly on the scheme (see Table I) to drastically change the character of membrane-mediated interaction. Therefore, the conclusion that the Hamm-Kozlov functional is modified only quantitatively and to an insignificant extent remains valid.

**Acknowledgments**
Grateful acknowledgement to M. A. Kalutskii for his valuable assistance in configuring the molecular dynamics parameters.

**Funding**
The work was supported by the Ministry of Science and Higher Education of the Russian Federation.

**References**
1. Nelson, L.D.; Cox, M.M. *Lehninger Principles of Biochemistry*; W. H. Freeman: New York, USA, 2004;
2. Yeagle, P.L. *The Membranes of Cells*; Academic Press: Oxford, UK, 2016; ISBN 978-0-12-800047-2.
3. Canham, P.B. The Minimum Energy of Bending as a Possible Explanation of the Biconcave Shape of the Human Red Blood Cell. *J. Theor. Biol.* **1970**, *26*, 61–81, doi:10.1016/S0022-5193(70)80032-7.
4. Helfrich, W. Elastic Properties of Lipid Bilayers: Theory and Possible Experiments. *Z. Naturforsch. C* **1973**, *28*, 693–703, doi:10.1515/znc-1973-11-1209.
5. Deuling, H.J.; Helfrich, W. Red Blood Cell Shapes as Explained on the Basis of Curvature Elasticity. *Biophys. J.* **1976**, *16*, 861–868, doi:10.1016/S0006-3495(76)85736-0.
6. Evans, E.; Rawicz, W. Entropy-Driven Tension and Bending Elasticity in Condensed-Fluid Membranes. *Phys. Rev. Lett.* **1990**, *64*, 2094, doi:10.1103/PhysRevLett.64.2094.
7. Bashkirov, P. V.; Kuzmin, P.I.; Lillo, J. V.; Frolov, V.A. Molecular Shape Solution for Mesoscopic Remodeling of Cellular Membranes. *Annu. Rev. Biophys.* **2022**, *51*, 473, doi:10.1146/annurev-biophys-011422-100054.

8. Helfrich, W. Blocked Lipid Exchange in Bilayers and Its Possible Influence on the Shape of Vesicles. *Zeitschrift für Naturforsch. C* **1974**, *29*, 510–515, doi:10.1515/znc-1974-9-1010.
9. Hamm, M.; Kozlov, M.M. Elastic Energy of Tilt and Bending of Fluid Membranes. *Eur. Phys. J. E* **2000**, *3*, 323–335, doi:10.1007/s101890070003.
10. Terzi, M.M.; Ergüder, M.F.; Deserno, M. A Consistent Quadratic Curvature-Tilt Theory for Fluid Lipid Membranes. *J. Chem. Phys.* **2019**, *151*, 164108, doi:10.1063/1.5119683.
11. Pinigin, K. V.; Kuzmin, P.I.; Akimov, S.A.; Galimzyanov, T.R. Additional Contributions to Elastic Energy of Lipid Membranes: Tilt-Curvature Coupling and Curvature Gradient. *Phys. Rev. E* **2020**, *102*, 042406, doi:10.1103/PhysRevE.102.042406.
12. Kollmitzer, B.; Heftberger, P.; Rappolt, M.; Pabst, G. Monolayer Spontaneous Curvature of Raft-Forming Membrane Lipids. *Soft Matter* **2013**, *9*, 10877–10884, doi:10.1039/C3SM51829A.
13. Campelo, F.; Arnarez, C.; Marrink, S.J.; Kozlov, M.M. Helfrich Model of Membrane Bending: From Gibbs Theory of Liquid Interfaces to Membranes as Thick Anisotropic Elastic Layers. *Adv. Colloid Interface Sci.* **2014**, *208*, 25–33, doi:10.1016/j.cis.2014.01.018.
14. Kalutskii, M.A.; Galimzyanov, T.R.; Pinigin, K. V. Determination of Elastic Parameters of Lipid Membranes from Simulation under Varied External Pressure. *Phys. Rev. E* **2023**, *107*, 024414, doi:10.1103/PhysRevE.107.024414.
15. Pinigin, K. V. Local Stress in Cylindrically Curved Lipid Membrane: Insights into Local versus Global Lateral Fluidity Models. *Biomolecules* **2024**, *14*, 1471, doi:10.3390/biom14111471.
16. May, S. Protein-Induced Bilayer Deformations: The Lipid Tilt Degree of Freedom. *Eur. Biophys. J.* **2000**, *29*, 17–28.
17. Kheyfets, B.; Mukhin, S.; Galimzyanov, T. Origin of Lipid Tilt in Flat Monolayers and Bilayers. *Phys. Rev. E* **2019**, *100*, 062405, doi:10.1103/PhysRevE.100.062405.
18. Bohinc, K.; Kralj-Iglič, V.; May, S. Interaction between Two Cylindrical Inclusions in a Symmetric Lipid Bilayer. *J. Chem. Phys.* **2003**, *119*, 7435–7444.
19. May, S.; Ben-Shaul, A. Molecular Theory of Lipid-Protein Interaction and the Lα-HII Transition. *Biophys. J.* **1999**, *76*, 751–767, doi:10.1016/S0006-3495(99)77241-3.
20. Müller, M.M.; Deserno, M.; Guven, J. Interface-Mediated Interactions between Particles: A Geometrical Approach. *Phys. Rev. E* **2005**, *72*, 061407, doi:10.1103/PhysRevE.72.061407.
21. Fournier, J.B. Coupling between Membrane Tilt-Difference and Dilation: A New "Ripple" Instability and Multiple Crystalline Inclusions Phases. *EPL* **1998**, *43*, 725, doi:10.1209/epl/i1998-00424-4.
22. Fournier, J.B. Microscopic Membrane Elasticity and Interactions among Membrane Inclusions: Interplay between the Shape, Dilation, Tilt and Tilt-Difference Modes. *Eur. Phys. J. B* **1999**, *11*, 261–272, doi:10.1007/BF03219168.
23. Galimzyanov, T.R.; Molotkovsky, R.J.; Bozdaganyan, M.E.; Cohen, F.S.; Pohl, P.; Akimov, S.A. Elastic Membrane Deformations Govern Interleaflet Coupling of Lipid-Ordered Domains. *Phys. Rev. Lett.* **2015**, *115*, 088101, doi:10.1103/PhysRevLett.115.088101.
24. Staneva, G.; Osipenko, D.S.; Galimzyanov, T.R.; Pavlov, K. V.; Akimov, S.A. Metabolic Precursor of Cholesterol Causes Formation of Chained Aggregates of Liquid-Ordered Domains. *Langmuir* **2016**, *32*, 1591–1600, doi:10.1021/acs.langmuir.5b03990.

25. Pinigin, K. V.; Kondrashov, O. V.; Jiménez-Munguía, I.; Alexandrova, V. V.; Batishchev, O. V.; Galimzyanov, T.R.; Akimov, S.A. Elastic Deformations Mediate Interaction of the Raft Boundary with Membrane Inclusions Leading to Their Effective Lateral Sorting. *Sci. Rep.* **2020**, *10*, 4087, doi:10.1038/s41598-020-61110-2.
26. Pinigin, K. V.; Galimzyanov, T.R.; Akimov, S.A. Amphipathic Peptides Impede Lipid Domain Fusion in Phase-Separated Membranes. *Membranes (Basel).* **2021**, *11*, 797, doi:10.3390/membranes11110797.
27. Pinigin, K. V.; Akimov, S.A. The Membrane-Mediated Interaction of Liquid-Ordered Lipid Domains in the Presence of Amphipathic Peptides. *Membranes (Basel).* **2023**, *13*, 816, doi:10.3390/membranes13100816.
28. Kondrashov, O.V.; Pinigin, K.V.; Akimov, S.A. Characteristic Lengths of Transmembrane Peptides Controlling Their Tilt and Lateral Distribution between Membrane Domains. *Phys. Rev. E* **2021**, *104*, 044411, doi:10.1103/PhysRevE.104.044411.
29. Campelo, F.; McMahon, H.T.; Kozlov, M.M. The Hydrophobic Insertion Mechanism of Membrane Curvature Generation by Proteins. *Biophys. J.* **2008**, *95*, 2325–2339, doi:10.1529/biophysj.108.133173.
30. Akimov, S.A.; Volynsky, P.E.; Galimzyanov, T.R.; Kuzmin, P.I.; Pavlov, K. V.; Batishchev, O. V. Pore Formation in Lipid Membrane I: Continuous Reversible Trajectory from Intact Bilayer through Hydrophobic Defect to Transversal Pore. *Sci. Rep.* **2017**, *7*, 12509, doi:10.1038/s41598-017-12127-7.
31. Akimov, S.A.; Volynsky, P.E.; Galimzyanov, T.R.; Kuzmin, P.I.; Pavlov, K. V.; Batishchev, O. V. Pore Formation in Lipid Membrane II: Energy Landscape under External Stress. *Sci. Rep.* **2017**, *7*, 12152, doi:10.1038/s41598-017-12749-x.
32. Kozlovsky, Y.; Kozlov, M.M. Stalk Model of Membrane Fusion: Solution of Energy Crisis. *Biophys. J.* **2002**, *82*, 882–895, doi:10.1016/S0006-3495(02)75450-7.
33. Kuzmin, P.I.; Zimmerberg, J.; Chizmadzhev, Y.A.; Cohen, F.S. A Quantitative Model for Membrane Fusion Based on Low-Energy Intermediates. *Proc. Natl. Acad. Sci.* **2001**, *98*, 7235–7240, doi:10.1073/pnas.121191898.
34. Bashkirov, P. V.; Akimov, S.A.; Evseev, A.I.; Schmid, S.L.; Zimmerberg, J.; Frolov, V.A. GTPase Cycle of Dynamin Is Coupled to Membrane Squeeze and Release, Leading to Spontaneous Fission. *Cell* **2008**, *135*, 1276–1286, doi:10.1016/j.cell.2008.11.028.
35. Shnyrova, A. V.; Bashkirov, P. V.; Akimov, S.A.; Pucadyil, T.J.; Zimmerberg, J.; Schmid, S.L.; Frolov, V.A. Geometric Catalysis of Membrane Fission Driven by Flexible Dynamin Rings. *Science.* **2013**, *339*, 1433–1436, doi:10.1126/science.1233920.
36. Ogden, R. *Non-Linear Elastic Deformations*; Dover Publications: Mineola, New York, 1997;
37. Braganza, L.F.; Worcester, D.L. Structural Changes in Lipid Bilayers and Biological Membranes Caused by Hydrostatic Pressure. *Biochemistry* **1986**, *25*, 7484–7488, doi:10.1021/bi00371a034.
38. Scarlata, S.F. Compression of Lipid Membranes as Observed at Varying Membrane Positions. *Biophys. J.* **1991**, *60*, 334–340, doi:10.1016/S0006-3495(91)82058-6.
39. Vennemann, N.; Lechner, M.D.; Henkel, T.; Knoll, W. Densitometric Characterization of the Main Phase Transition of Dimyristoyl-Phosphatidylcholine between 0.1 and 40 MPa. *Berichte der Bunsengesellschaft für Phys. Chemie* **1986**, *90*, 888–891, doi:10.1002/bbpc.19860901011.
40. Terzi, M.M.; Deserno, M.; Nagle, J.F. Mechanical Properties of Lipid Bilayers: A Note on the Poisson Ratio. *Soft Matter* **2019**, *15*, 9085–9092,

doi:10.1039/C9SM01290G.
41. Berger, O.; Edholm, O.; Jähnig, F. Molecular Dynamics Simulations of a Fluid Bilayer of Dipalmitoylphosphatidylcholine at Full Hydration, Constant Pressure, and Constant Temperature. *Biophys. J.* **1997**, *72*, 2002–2013, doi:10.1016/S0006-3495(97)78845-3.
42. Venable, R.M.; Skibinsky, A.; Pastor, R.W. Constant Surface Tension Molecular Dynamics Simulations of Lipid Bilayers with Trehalose. *Mol. Simul.* **2006**, *32*, 849–855, doi:10.1080/08927020600615018.
43. Truesdell, C.; Noll, W. *The Non-Linear Field Theories of Mechanics*; 3rd ed.; Springer-Verlag: Berlin, 2004;
44. Gurtin, M.E.; Fried, E.; Anand, L. *The Mechanics and Thermodynamics of Continua*; Cambridge University Press: Cambridge, 2010;
45. Horgan, C.O.; Murphy, J.G. Poynting and Reverse Poynting Effects in Soft Materials. *Soft Matter* **2017**, *13*, 4916–4923, doi:10.1039/C7SM00992E.
46. Zurlo, G.; Blackwell, J.; Colgan, N.; Destrade, M. The Poynting Effect. *Am. J. Phys.* **2020**, *88*, 1036–1040, doi:10.1119/10.0001997.
47. Marrink, S.J.; Risselada, H.J.; Yefimov, S.; Tieleman, D.P.; De Vries, A.H. The MARTINI Force Field: Coarse Grained Model for Biomolecular Simulations. *J. Phys. Chem. B* **2007**, *111*, 7812–7824, doi:10.1021/jp071097f.
48. Arnarez, C.; Uusitalo, J.J.; Masman, M.F.; Ingólfsson, H.I.; De Jong, D.H.; Melo, M.N.; Periole, X.; de Vries, A.H.; Marrink, S.J. Dry Martini, a Coarse-Grained Force Field for Lipid Membrane Simulations with Implicit Solvent. *J. Chem. Theory Comput.* **2015**, *11*, 260–275, doi:10.1021/ct500477k.
49. Lorent, J.H.; Levental, K.R.; Ganesan, L.; Rivera-Longsworth, G.; Sezgin, E.; Doktorova, M.; Lyman, E.; Levental, I. Plasma Membranes Are Asymmetric in Lipid Unsaturation, Packing and Protein Shape. *Nat. Chem. Biol.* **2020**, *16*, 644–652, doi:10.1038/s41589-020-0564-3.
50. Luchini, A.; Vitiello, G. Mimicking the Mammalian Plasma Membrane: An Overview of Lipid Membrane Models for Biophysical Studies. *Biomimetics* **2020**, *6*, 3, doi:10.3390/biomimetics6010003.
51. Pluhackova, K.; Kirsch, S.A.; Han, J.; Sun, L.; Jiang, Z.; Unruh, T.; Böckmann, R.A. A Critical Comparison of Biomembrane Force Fields: Structure and Dynamics of Model DMPC, POPC, and POPE Bilayers. *J. Phys. Chem. B* **2016**, *120*, 3888–3903, doi:10.1021/acs.jpcb.6b01870.
52. Piskulich, Z.A.; Cui, Q. Machine Learning-Assisted Phase Transition Temperatures from Generalized Replica Exchange Simulations of Dry Martini Lipid Bilayers. *J. Phys. Chem. Lett.* **2022**, *13*, 6481–6486, doi:10.1021/acs.jpclett.2c01654.
53. Van Der Spoel, D.; Lindahl, E.; Hess, B.; Groenhof, G.; Mark, A.E.; Berendsen, H.J. GROMACS: Fast, Flexible, and Free. *J. Comput. Chem.* **2005**, *26*, 1701–1718, doi:10.1002/jcc.20291.
54. Abraham, M.J.; Murtola, T.; Schulz, R.; Páll, S.; Smith, J.C.; Hess, B.; Lindahl, E. GROMACS: High Performance Molecular Simulations through Multi-Level Parallelism from Laptops to Supercomputers. *SoftwareX* **2015**, *1*, 19–25, doi:10.1016/j.softx.2015.06.001.
55. Goga, N.; Rzepiela, A.J.; De Vries, A.H.; Marrink, S.J.; Berendsen, H.J.C. Efficient Algorithms for Langevin and DPD Dynamics. *J. Chem. Theory Comput.* **2012**, *8*, 3637–3649, doi:10.1021/ct3000876.
56. Verlet, L. Computer" Experiments" on Classical Fluids. I. Thermodynamical Properties of Lennard-Jones Molecules. *Phys. Rev.* **1967**, *159*, 98, doi:10.1103/PhysRev.159.98.
57. Berendsen, H.J.C.; Postma, J.P.M.; van Gunsteren, W.F.; DiNola, A.; Haak, J.R.

Molecular Dynamics with Coupling to an External Bath. *J. Chem. Phys.* **1984**, *81*, 3684–3690, doi:10.1063/1.448118.
58. Fiorin, G.; Klein, M.L.; Hénin, J. Using Collective Variables to Drive Molecular Dynamics Simulations. *Mol. Phys.* **2013**, *111*, 3345–3362, doi:10.1080/00268976.2013.813594.
59. Vanegas, J.M.; Torres-Sánchez, A.; Arroyo, M. Importance of Force Decomposition for Local Stress Calculations in Biomembrane Molecular Simulations. *J. Chem. Theory Comput.* **2014**, *10*, 691–702, doi:10.1021/ct4008926.
60. Torres-Sánchez, A.; Vanegas, J.M.; Arroyo, M. Examining the Mechanical Equilibrium of Microscopic Stresses in Molecular Simulations. *Phys. Rev. Lett.* **2015**, *114*, 258102, doi:10.1103/PhysRevLett.114.258102.
61. Torres-Sánchez, A.; Vanegas, J.M.; Arroyo, M. Geometric Derivation of the Microscopic Stress: A Covariant Central Force Decomposition. *J. Mech. Phys. Solids* **2016**, *93*, 224–239, doi:10.1016/j.jmps.2016.03.006.
62. Ollila, O.S.; Risselada, H.J.; Louhivuori, M.; Lindahl, E.; Vattulainen, I.; Marrink, S.J. 3D Pressure Field in Lipid Membranes and Membrane-Protein Complexes. *Phys. Rev. Lett.* **2009**, *102*, 078101, doi:10.1103/PhysRevLett.102.078101.
63. Irving, J.H.; Kirkwood, J.G. The Statistical Mechanical Theory of Transport Processes. IV. The Equations of Hydrodynamics. *J. Chem. Phys.* **1950**, *18*, 817–829, doi:10.1063/1.1747782.
64. Lehoucq, R.B.; Von Lilienfeld-Toal, A. Translation of Walter Noll's "Derivation of the Fundamental Equations of Continuum Thermodynamics from Statistical Mechanics." *J. Elast.* **2010**, *100*, 5–24, doi:10.1007/s10659-010-9246-9.
65. Michaud-Agrawal, N.; Denning, E.J.; Woolf, T.B.; Beckstein, O. MDAnalysis: A Toolkit for the Analysis of Molecular Dynamics Simulations. *J. Comput. Chem.* **2011**, *32*, 2319–2327, doi:10.1002/jcc.21787.
66. Gowers, R.J.; Linke, M.; Barnoud, J.; Reddy, T.J.E.; Melo, M.N.; Seyler, S.L.; Dotson, D.L.; Domanski, J.; Buchoux, S.; Kenney, I.M.; et al. MDAnalysis: A Python Package for the Rapid Analysis of Molecular Dynamics Simulations. In Proceedings of the 15th Python in Science Conference; Benthall, S., Rostrup, S., Eds.; Austin, TX, 2016; pp. 98–105.
67. May, E.R.; Narang, A.; Kopelevich, D.I. Role of Molecular Tilt in Thermal Fluctuations of Lipid Membranes. *Phys. Rev. E* **2007**, *76*, 021913, doi:10.1103/PhysRevE.76.021913.
68. Watson, M.C.; Penev, E.S.; Welch, P.M.; Brown, F.L. Thermal Fluctuations in Shape, Thickness, and Molecular Orientation in Lipid Bilayers. *J. Chem. Phys.* **2011**, *135*, 244701, doi:10.1063/1.3660673.
69. Watson, M.C.; Brandt, E.G.; Welch, P.M.; Brown, F.L. Determining Biomembrane Bending Rigidities from Simulations of Modest Size. *Phys. Rev. Lett.* **2012**, *109*, 028102, doi:10.1103/PhysRevLett.109.028102.
70. Levine, Z.A.; Venable, R.M.; Watson, M.C.; Lerner, M.G.; Shea, J.E.; Pastor, R.W.; Brown, F.L. Determination of Biomembrane Bending Moduli in Fully Atomistic Simulations. *J. Am. Chem. Soc.* **2014**, *136*, 13582–13585, doi:10.1021/ja507910r.
71. Ergüder, M.F.; Deserno, M. Identifying Systematic Errors in a Power Spectral Analysis of Simulated Lipid Membranes. *J. Chem. Phys.* **2021**, *154*, 214103, doi:10.1063/5.0049448.
72. Deserno, M. Membrane Elasticity and Mediated Interactions in Continuum Theory: A Differential Geometric Approach. In *Biomembrane frontiers: nanostructures, models, and the design of life*; Faller, R., Longo, M., Risbud, S., Jue, T., Eds.; Springer: Dordrecht, 2009; pp. 41–74.

73. Terzi, M.M.; Deserno, M. Novel Tilt-Curvature Coupling in Lipid Membranes. *J. Chem. Phys.* **2017**, *147*, 084702, doi:10.1063/1.4990404.
74. Akimov, S.A.; Aleksandrova, V. V.; Galimzyanov, T.R.; Bashkirov, P. V.; Batishchev, O. V. Interaction of Amphipathic Peptides Mediated by Elastic Membrane Deformations. *Biol. Membr.* **2017**, *34*, 162–173, doi:10.7868/S0233475517030033.
75. Kondrashov, O. V.; Galimzyanov, T.R.; Jiménez-Munguía, I.; Batishchev, O. V.; Akimov, S.A. Membrane-Mediated Interaction of Amphipathic Peptides Can Be Described by a One-Dimensional Approach. *Phys. Rev. E* **2019**, *99*, 022401, doi:10.1103/PhysRevE.99.022401.